\documentstyle[epsfig,rotate,12pt]{article}
\newlength{\dinwidth}                       
\newlength{\dinmargin}                      
\def\lsim{\mathrel{\rlap{\lower4pt\hbox{\hskip1pt$\sim$}}
    \raise1pt\hbox{$<$}}}                % less than or approx. symbol
\def\gsim{\mathrel{\rlap{\lower4pt\hbox{\hskip1pt$\sim$}}
    \raise1pt\hbox{$>$}}}                % greater than or approx. symbol
\def\sss{\scriptscriptstyle}
\def\barp{{\raise.35ex\hbox{${\sss (}$}}---{\raise.35ex\hbox{${\sss )}$}}}
\def\bdbarp{\hbox{$B^0$\kern-1.4em\raise1.4ex\hbox{\barp}}}
\def\bsbarp{\hbox{$B_s$\kern-1.4em\raise1.4ex\hbox{\barp}}}
\def\dbarp{\hbox{$D$\kern-1.1em\raise1.4ex\hbox{\barp}}}

\newcommand{\xs}{x_s}
\newcommand{\bd}{B_d^0}
\newcommand{\bdb}{\overline{B_d^0}}
\newcommand{\bs}{B_s^0}
\newcommand{\bsb}{\overline{B_s^0}}

\newcommand{\beq}{\begin{equation}}
\newcommand{\eeq}{\end{equation}}
\newcommand{\absvcb}{\vert V_{cb}\vert}
\newcommand{\absvub}{\vert V_{ub}\vert}
\newcommand{\absvtd}{\vert V_{td}\vert}
\newcommand{\absvts}{\vert V_{ts}\vert}
\newcommand{\abseps}{\vert\epsilon_K\vert}

\newcommand{\fbb}{f^2_{B_d}\hat{B}_{B_d}}
\newcommand{\fbbs}{f^2_{B_s}\hat{B}_{B_s}}
\newcommand{\fbd}{f_{B_d}}

\def\rly#1{\mathrel{\raise.3ex\hbox{$#1$\kern-.75em\lower1ex\hbox{$\sim$}}}}
\def\lsim{\rly<}

\newread\epsffilein % file to \read
\newif\ifepsffileok % continue looking for the bounding box?
\newif\ifepsfbbfound % success?
\newif\ifepsfverbose % report what you're making?
\newdimen\epsfxsize % horizontal size after scaling
\newdimen\epsfysize % vertical size after scaling
\newdimen\epsftsize % horizontal size before scaling
\newdimen\epsfrsize % vertical size before scaling
\newdimen\epsftmp % register for arithmetic manipulation
\newdimen\pspoints % conversion factor
\def\epsfbox#1{\global\def\epsfllx{72}\global\def\epsflly{72}%
 \global\def\epsfurx{540}\global\def\epsfury{720}%
 \def\lbracket{[}\def\testit{#1}\ifx\testit\lbracket
 \let\next=\epsfgetlitbb\else\let\next=\epsfnormal\fi\next{#1}}%
\def\epsfgetlitbb#1#2 #3 #4 #5]#6{\epsfgrab #2 #3 #4 #5 .\\%
 \epsfsetgraph{#6}}%
\def\epsfnormal#1{\epsfgetbb{#1}\epsfsetgraph{#1}}%
\def\epsfgetbb#1{%
\openin\epsffilein=#1
\ifeof\epsffilein\errmessage{I couldn't open #1, will ignore it}\else
 {\epsffileoktrue \chardef\other=12
 \def\do##1{\catcode`##1=\other}\dospecials \catcode`\ =10
 \loop
 \read\epsffilein to \epsffileline
 \ifeof\epsffilein\epsffileokfalse\else
 \expandafter\epsfaux\epsffileline:. \\%
 \fi
 \ifepsffileok\repeat
 \ifepsfbbfound\else
 \ifepsfverbose\message{No bounding box comment in #1; using defaults}\fi\fi
 }\closein\epsffilein\fi}%
\def\epsfclipstring{}% do we clip or not? If so,
\def\epsfsetgraph#1{%
 \epsfrsize=\epsfury\pspoints
 \advance\epsfrsize by-\epsflly\pspoints
 \epsftsize=\epsfurx\pspoints
 \advance\epsftsize by-\epsfllx\pspoints
 \epsfxsize\epsfsize\epsftsize\epsfrsize
 \ifnum\epsfxsize=0 \ifnum\epsfysize=0
 \epsfxsize=\epsftsize \epsfysize=\epsfrsize
 \epsfrsize=0pt
 \else\epsftmp=\epsftsize \divide\epsftmp\epsfrsize
 \epsfxsize=\epsfysize \multiply\epsfxsize\epsftmp
 \multiply\epsftmp\epsfrsize \advance\epsftsize-\epsftmp
 \epsftmp=\epsfysize
 \loop \advance\epsftsize\epsftsize \divide\epsftmp 2
 \ifnum\epsftmp>0
 \ifnum\epsftsize<\epsfrsize\else
 \advance\epsftsize-\epsfrsize \advance\epsfxsize\epsftmp \fi
 \repeat
 \epsfrsize=0pt
 \fi
 \else \ifnum\epsfysize=0
 \epsftmp=\epsfrsize \divide\epsftmp\epsftsize
 \epsfysize=\epsfxsize \multiply\epsfysize\epsftmp
 \multiply\epsftmp\epsftsize \advance\epsfrsize-\epsftmp
 \epsftmp=\epsfxsize
 \loop \advance\epsfrsize\epsfrsize \divide\epsftmp 2
 \ifnum\epsftmp>0
 \ifnum\epsfrsize<\epsftsize\else
 \advance\epsfrsize-\epsftsize \advance\epsfysize\epsftmp \fi
 \repeat
 \epsfrsize=0pt
 \else
 \epsfrsize=\epsfysize
 \fi
 \fi
 \ifepsfverbose\message{#1: width=\the\epsfxsize, height=\the\epsfysize}\fi
 \epsftmp=10\epsfxsize \divide\epsftmp\pspoints
 \vbox to\epsfysize{\vfil\hbox to\epsfxsize{%
 \ifnum\epsfrsize=0\relax
 \includegraphics{#1}%
 \else
 \epsfrsize=10\epsfysize \divide\epsfrsize\pspoints
 \includegraphics{#1}%
 \fi
 \hfil}}%
\global\epsfxsize=0pt\global\epsfysize=0pt}%
 {\catcode`\%=12 \global\let\epsfpercent=%\global\def\epsfbblit{%BoundingBox}}%
\long\def\epsfaux#1#2:#3\\{\ifx#1\epsfpercent
 \def\testit{#2}\ifx\testit\epsfbblit
 \epsfgrab #3 . . . \\%
 \epsffileokfalse
 \global\epsfbbfoundtrue
 \fi\else\ifx#1\par\else\epsffileokfalse\fi\fi}%
\def\epsfempty{}%
\def\epsfgrab #1 #2 #3 #4 #5\\{%
\global\def\epsfllx{#1}\ifx\epsfllx\epsfempty
 \epsfgrab #2 #3 #4 #5 .\\\else
 \global\def\epsflly{#2}%
 \global\def\epsfurx{#3}\global\def\epsfury{#4}\fi}%
\def\epsfsize#1#2{\epsfxsize}
\def\g{\gamma}

\def\mt{m_t}

\def\be{\begin{equation}}
\def\ee{\end{equation}}

\newcommand{\delmd}{\Delta M_d}
\newcommand{\delms}{\Delta M_s}

\newcommand{\bdbdbar}{$B_d^0$-${\overline{B_d^0}}$}
\newcommand{\bsbsbar}{$B_s^0$-${\overline{B_s^0}}$}

\def\bxsll{$B \rightarrow X_s \ell^+ \ell^- $ }
\def\bxsee{B \rightarrow X_s e^+ e^-  }
\def\bxsmm{B \rightarrow X_s \mu^+ \mu^-  }
\def\bxstt{B \rightarrow X_s \tau^+ \tau^- }

\def\bxsg{$B \rightarrow X_s \gamma $ }

\newcommand{\ra}{\rightarrow}
\newcommand{\bgamaxs}{$B \to X _{s} + \gamma$}

\newcommand{\BGAMAXS}{B \ra X _{s} + \gamma}
\newcommand{\BGAMAXD}{B \ra X _{d} + \gamma}
\newcommand{\BBGAMAXS}{{\cal B}(B \ra  X _{s} + \gamma)}
\newcommand{\BBGAMAXD}{{\cal B}(B \ra  X _{d} + \gamma)}

\newcommand{\BGAMAKSTAR}{B \ra  K^{\star} + \gamma}

\newcommand{\bgamaxd}{$B \to X _{d} + \gamma$}

\def\Vcbabs{\vert V_{cb} \vert}

\def\Vubabs{\vert V_{ub}\vert}

\def\Vtdabs{\vert V_{td} \vert}

\def\Vtsabs{\vert V_{ts} \vert}

\textheight 23.0 true cm
\begin{document}
\vspace*{1cm}
\begin{center}  \begin{Large} \begin{bf}
Heavy Flavour Decays - Introduction and Summary\\
  \end{bf}  \end{Large}
  \vspace*{5mm}
  \begin{large}
A. Ali$^a$ and H.Schr\"oder$^a$\\
  \end{large}   
\end{center}
$^a$ Deutsches~Elektronen-Synchrotron~DESY,
     Notkestrasse~85,~D-22603~Hamburg,~FRG\\
\begin{quotation}
\noindent
{\bf Abstract:}
We review some selected topics in the 
decays of heavy flavours, beauty and
charm, which are of principal interest at HERA-B and HERA.
The topics in $B$ physics include: an update on the quark mixing matrix
and CP violating phases, issues bearing on an improved resolution on 
the CP-violating phase $\Delta (\sin 2 \beta)$,
prospects of measuring  radiative and semileptonic rare $B$ decays,
the $\bs$ - $\bsb$ mixing ratio $x_s$, improved
measurements of the $\bd$ - $\bdb$ mixing ratio $x_d$ and the $B$-hadron 
lifetimes, in particular $\tau(\Lambda_b)$.
 In the charm sector, we have focussed on rare decays and 
$D^0$- $\overline{D^0}$ mixing, whose measurements will signal physics
beyond the standard model.
\end{quotation}
%%%%%%%%%%%%%%%%%%%%%% SECTION 1 %%%%%%%%%%%%%%%%%%%%%%%%%%%%%%%%
\section{Introduction}
\begin{figure}
\vskip -1.0truein
\centerline{\epsfxsize 3.5 truein \epsfbox {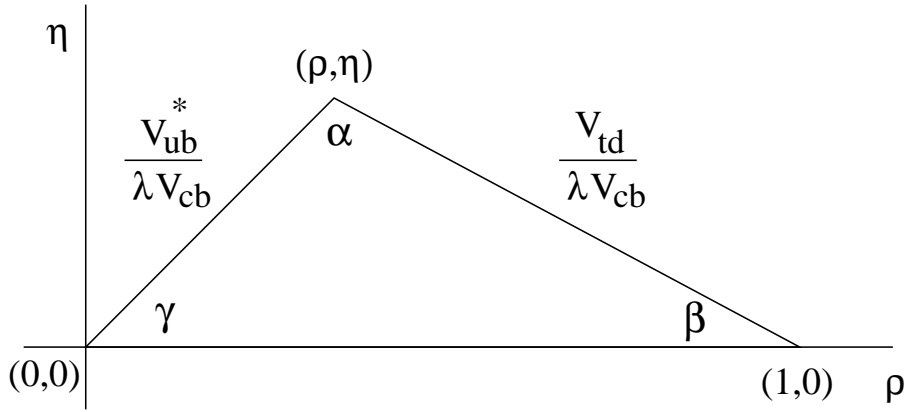}}
\vskip -1.2truein
\caption{The unitarity triangle. The angles $\alpha$, $\beta$ and $\gamma$
can be measured via CP violation in the $B$ system.}
\label{triangle}
\end{figure}

    The origin of CP violation, even 32 years after its discovery by
Christenson et al. in neutral $K$ decays
\cite{CPV64}, remains a puzzle.  So far the ratio $\epsilon_K$ is the
only precisely determined CP violating quantity in particle physics
\cite{PDG96}. In the 
standard model (SM), the
couplings of the charged vector bosons $W^\pm$ with the quark-antiquark
pairs are complex, which for three generations lead to a complex phase
in the quark mixing matrix - the
 Cabibbo-Kobayashi-Maskawa (CKM) matrix \cite{CKM}. The quantity
$\epsilon_K$ in the SM measures essentially this complex phase.
However, this hypothesis remains to be tested by independent 
measurements of other CP-violating quantities. In neutral $B$-meson decays,
the angles in the CKM-unitarity triangle shown in Fig.~\ref{triangle}
are characteristic measures of
CP violation, as the CP violating asymmetries in the decay rates 
of a $B$-hadron into specific modes and their CP-conjugates can be 
related to these angles. The primary aim of the HERA-B
experiment is to measure the CP-violating asymmetry in $B$ decays
related to the phase $\sin (2 \beta)$.  

 A related and equally important goal of the
HERA-B physics programme is to quantitatively test the unitarity of the
CKM matrix, in which apart from the improved measurements of the matrix 
elements $\Vcbabs$ and $\Vubabs$, the matrix element $\Vtdabs$ plays a
central role (see Fig.~\ref{triangle}). This matrix element can, in 
principle, be determined in a number of $B$ and $K$ decays \cite{ALI96,BBL95}.
At HERA-B, this would require either measuring
the mass difference between the two eigenstates of the \bsbsbar ~system,  
$\Delta M_s$, (equivalently the ratio of the mass difference to the averaged
decay width $x_s = \Delta M_s/\Gamma_s$), which can then be compared
with the already well-measured quantity $\delmd$ \cite{Gibbons96} to 
extract the ratio $\Vtdabs/\Vtsabs$, or the branching ratios of at least 
one of 
the CKM-suppressed rare decays, such as $B^{0} \to \rho^{0} + \gamma$,
which would yield $\Vtdabs/\Vcbabs$.
These measurements are also very challenging; apart from efficient
triggers and good vertex resolution, 
both high luminosity and higher proton beam energy will be big assets here.
 The unitarity of the CKM matrix will, of course, be very precisely
tested by the measurements of all three angles in the unitarity triangle 
shown in Fig.~\ref{triangle}. This, however, is an ambitious programme,
which may or may not materialize in three years of HERA-B running and
may require a post-HERA-B facility such as the LHC. 
Hence, in this workshop, we have concentrated on $\sin (2 \beta)$, $x_s$,
and rare $B$ decays.

% This is Figure 1

 Concerning charm physics at HERA and HERA-B, which will 
undoubtedly contribute to our 
understanding of the dynamics of heavy flavour production in QCD 
(reviewed here
by Eichler and Frixione \cite{EFHERA96}), the principal interest is in
attaining improved experimental sensitivities in searches for
rare decays, $D^{0}$ - $\overline{D^{0}}$ mixing and CP violation. As opposed
to the FCNC phenomena in $B$ decays, SM predicts tiny FCNC decay rates,
mixing frequency, and CP asymmetries in the charm sector. This reflects 
the observed pattern of quark masses and mixings and the
built-in GIM-mechanism \cite{GIM} in the SM. Long-distance (LD)
effects may increase some of the transition rates but
these enhancements in all realistic estimates remain modest; FCNC-related 
phenomena in the charm sector remain unmeasurable in SM for all practical 
purposes.
 Hence, finding a positive signal in any 
of the rare decays such as $D^0 \to \ell^+ \ell^-$, $D^{0} \to \gamma 
\gamma$, $D^{0}$ - $\overline{D^{0}}$ mixing or CP violation in any
charmed hadron decay mode
will unambiguously signal physics beyond the SM. As argued quantitatively 
in these proceedings by Grab \cite{Grab96} and Tsipolitis 
\cite{Tsipolitis96}, counting rate is the decisive parameter in such
searches and an increase in HERA-luminosity will be very welcome in 
the $ep$ mode. At HERA-B, the anticipated charmed hadron production 
rate is already very high. Here, one has to develop an efficient trigger
to make use of this high yield and do competitive physics.        

%%%%%%%%%%%%%%%%%%%%%% SECTION 2 %%%%%%%%%%%%%%%%%%%%%%%%%%%%%%%%
\section{Flavour Mixings and CP Violation in the SM and at HERA-B}
The profile of the CKM
matrix \cite{CKM} is updated by Ali and London in \cite{AL96}. 
 In particular, they focussed
on the CKM unitarity triangle and CP asymmetries in $B$ decays, which
are the principal objects of interest in experiments at present and 
forthcoming $B$ facilities, in particular HERA-B.

\subsection{Present profile of the CKM unitarity triangle}

In updating the CKM matrix elements the  Wolfenstein parametrization 
\cite{Wolfenstein} has been used which follows from the observation that
the elements of this matrix exhibit a hierarchy in terms of $\lambda$, the
Cabibbo angle. In this parametrization the CKM matrix can be written
approximately as
\beq
V_{CKM} \simeq \left(\matrix{
 1-{1\over 2}\lambda^2 & \lambda
 & A\lambda^3 \left( \rho - i\eta \right) \cr
 -\lambda ( 1 + i A^2 \lambda^4 \eta )
& 1-{1\over 2}\lambda^2 & A\lambda^2 \cr
 A\lambda^3\left(1 - \rho - i \eta\right) & -A\lambda^2 & 1 \cr}\right)~.
\label{CKM}
\eeq
The allowed region in $\rho$-$\eta$ space can be displayed quite elegantly
using the so-called unitarity triangle. The unitarity of the CKM matrix
leads to the following relation:
\beq
V_{ud} V_{ub}^* + V_{cd} V_{cb}^* + V_{td} V_{tb}^* = 0~.
\eeq
Using the form of the CKM matrix in Eq.~\ref{CKM}, this can be recast as
\beq
\label{trianglerel}
\frac{V_{ub}^*}{\lambda V_{cb}} + \frac{V_{td}}{\lambda V_{cb}} = 1~,
\eeq
which is a triangle relation in the complex plane (i.e.\ $\rho$-$\eta$
space), illustrated in Fig.~\ref{triangle}. Thus, allowed values of $\rho$
and $\eta$ translate into allowed shapes of the unitarity triangle.

\begin{figure}
\vskip -1.0truein
\centerline{\epsfxsize 3.5 truein \epsfbox {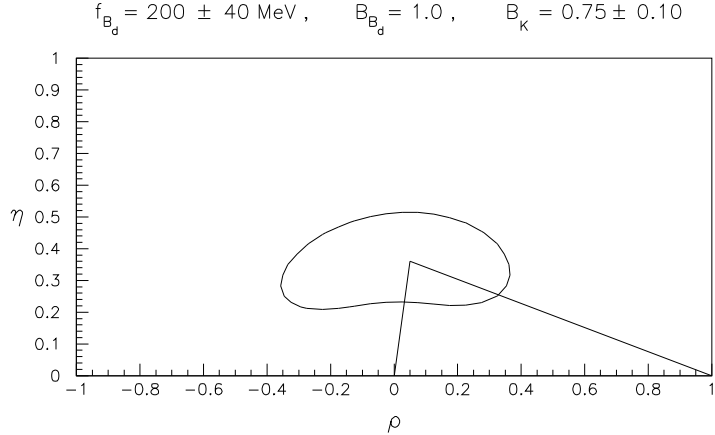}}
\vskip -1.4truein
\caption{Allowed region in $\rho$-$\eta$ space, from a simultaneous fit to
both the experimental and theoretical quantities given in
 eqs.~(\protect{\ref{alambda}}) and (\protect{\ref{ckmfiteqns}}).
The theoretical errors are treated as Gaussian for
this fit. The solid line represents the region with $\chi^2=\chi_{min}^2+6$
corresponding to the 95\% C.L.\ region. The triangle shows the best fit.
(From \protect\cite{AL96}.)}
\label{rhoeta2}
\end{figure}

The present status of the CKM-Wolfenstein parameters
$\lambda$ and  $A$ is as follows \cite{PDG96,Gibbons96}:
\begin{eqnarray}
\vert V_{us}\vert&=&\lambda=0.2205\pm 0.0018~, \nonumber\\
 \vert V_{cb} \vert &=& 0.0393 \pm 0.0028~ \Longrightarrow A = 0.81 \pm 
0.058~,
\label{alambda} 
\end{eqnarray}
The other two parameters $\rho$ and $\eta$ (the all important complex phase)
are determined at present through the measurements of $\absvub/\absvcb$,
$\delmd$, the \bdbdbar -mixing induced mass difference, and $\abseps$, the
CP-violating parameter in $K$ decays. The present experimental input
can be summarized as \cite{AL96}:
\begin{eqnarray}
\sqrt{\rho^2+\eta^2} &=& 0.363 \pm 0.073 ~~~\mbox{(from
$|V_{ub}/V_{cb}|=0.08 \pm 20\%$),} \nonumber \\
(\fbd \sqrt{\hat{B}_{B_d}}/\mbox{1 GeV}) \sqrt{(1-\rho)^2 + \eta^2} &=&
0.202 \pm 0.017 ~~~\mbox{(from $\delmd=0.464 \pm 0.018 ~(ps)^{-1}$),} 
\nonumber \\
\hat{B}_K \eta [ 0.93 + (2.08 \pm 0.34) (1-\rho)] &=& (0.79 \pm 0.11)
~~\mbox{(from $\abseps=(2.280 \pm 0.013)\times 10^{-3}$),}
\label{ckmfiteqns}
\end{eqnarray}
The errors of the last two lines include the small experimental errors on
$\delmd$ (3.9\%) and $\abseps$ (0.6\%), as well as the larger errors on
$m_t^2$ (11\%) and $A^2$ (14\%).
In \cite{AL96}, two types of CKM fits have been considered.
\begin{itemize}
\item
Fit 1: the ``experimental fit.'' Here, only the experimentally measured
numbers are used as inputs to the fit with Gaussian errors; the coupling
constants $f_{B_d} \sqrt{\hat{B}_{B_d}}$ and $\hat{B}_K$ are given fixed
values.
\item
Fit 2: the ``combined fit.'' Here, both the experimental and theoretical
numbers (indicated on Fig.~\ref{rhoeta2})
 are used as inputs assuming Gaussian errors for the theoretical
quantities. 
\end{itemize}
These two methods provide very similar results and we
focus here on the ``combined fit" (Fit 2), which  
is shown in terms of the allowed CKM triangle 
in Fig.~\ref{rhoeta2}. As is clear from this figure, the allowed region is
still rather large at present. However, present data and theory do
restrict the parameters $\rho$ and $\eta$ to lie in the following
range:
\begin{eqnarray}
 0.20 &\leq & \eta \leq 0.52 , \nonumber \\
 -0.35 &\leq & \rho \leq 0.35 ~.
\label{rhoetarange}
\end{eqnarray}
The preferred values obtained from the ``combined fit" are
\beq
(\rho,\eta) = (0.05,0.36) ~~~(\mbox{with}~\chi^2 = 6.6\times 10^{-3})~,
\label{bestrhoeta}
\eeq
which gives rise to an almost right-angled unitarity triangle, with the
angle $\gamma$ being close to $90$ degrees.
 
% This is Figure 2

\subsection{CP Violation in the $B$ System}

It is expected in the SM that the $B$ system will exhibit large CP-violating 
effects,
characterized by nonzero values of the angles $\alpha$, $\beta$ and
$\gamma$ in the unitarity triangle (Fig.~\ref{triangle}).
The most promising method to measure CP violation is to look for an
asymmetry between $\Gamma(B^0\to f)$ and $\Gamma({\overline{B^0}}\to f)$,
where $f$ is a CP eigenstate. If only one weak amplitude contributes to the
decay, the CKM phases can be extracted cleanly (i.e.\ with no hadronic
uncertainties). Thus, $\sin 2\alpha$, $\sin 2\beta$ and $\sin 2\gamma$ can
in principle be measured in $\bdbarp \to \pi^+ \pi^-$, $\bdbarp\to J/\psi
K_s$ and $\bsbarp\to\rho K_s$, respectively.
Penguin diagrams will, in general, introduce some hadronic
uncertainty into an otherwise clean measurement of the CKM phases. In the
case of $\bdbarp\to J/\psi K_s$, the penguins do not cause any problems,
since the weak phase of the penguin is the same as that of the tree
contribution. Thus, the CP asymmetry in this decay still measures $\sin
2\beta$. This augers well as measuring this asymmetry is the primary goal of
the HERA-B experiment.

 For $\bdbarp \to \pi^+ \pi^-$, however, although the penguin is
expected to be small with respect to the tree diagram, it will still
introduce a theoretical uncertainty into the extraction of $\alpha$. This
uncertainty can, in principle, be removed by the use of an isospin analysis,
which requires the measurement of the rates for
$B^+\to\pi^+\pi^0$, $B^0\to\pi^+\pi^-$ and $B^0\to\pi^0\pi^0$, as well as
their CP-conjugate counterparts. Help will come here from $e^+ e^-$
experiments which are the only ones which can measure the $B^0\to\pi^0\pi^0$ 
mode. Still, this isospin program is ambitious experimentally.
If it cannot be carried out, the error induced on $\sin 2\alpha$ is of order
$|P/T|$, where $P$ ($T$) represents the penguin (tree) diagram. The 
ratio $|P/T|$ is difficult to estimate since it is dominated by hadronic
physics. It is $\bsbarp\to\rho K_S$ which is most affected by penguins. 
In fact, the penguin contribution is probably larger in this process than 
the tree contribution. Other methods to measure $\gamma$ have been 
devised, not involving CP-eigenstate final states, and are reviewed in
\cite{AL96}.

\begin{figure}
\vskip -1.0truein
\centerline{\epsfxsize 3.5 truein \epsfbox {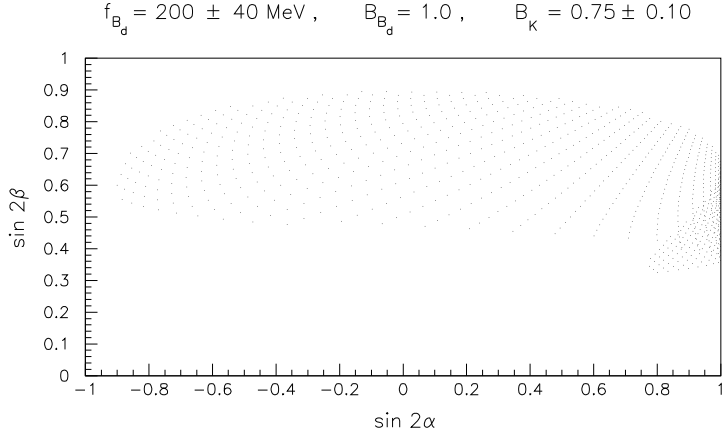}}
\vskip -1.4truein
\caption{Allowed region of the CP-violating quantities $\sin 2\alpha$ and
$\sin 2\beta$ resulting from the ``combined fit" of the data for the ranges
for $\fbd\protect\sqrt{\hat{B}_{B_d}} $ and $\hat{B}_K$ given in the text.
(From \protect\cite{AL96}.)}
\label{alphabeta2}
\end{figure}

The CP-violating asymmetries,
 parametrized by $\sin
2\alpha$, $\sin 2\beta$ and and $\sin^2 \gamma$ ,
 can be expressed straightforwardly in terms
of the CKM parameters $\rho$ and $\eta$. The 95\% C.L.\ constraints on
$\rho$ and $\eta$ found previously can be used to predict the ranges of
$\sin 2\alpha$, $\sin 2\beta$ and $\sin^2 \gamma$ allowed in the standard
model.  The ranges for the CP-violating rate asymmetries
 are determined at 95\% C.L. to be \cite{AL96}:
\begin{eqnarray}
&~& -0.90 \leq \sin 2\alpha \le 1.0~, \nonumber \\
&~& 0.32 \leq \sin 2\beta \le 0.94~, \\
&~& 0.34 \leq \sin^2 \gamma \le 1.0~. \nonumber
\end{eqnarray}
It is  assumed that the angle $\beta$ is measured in
$\bdbarp\to J/\psi K_s$, and an extra minus sign
due to the CP of the final state has been included.
Since the CP asymmetries all depend on $\rho$ and $\eta$, these ranges for
$\sin 2\alpha$, $\sin 2\beta$ and $\sin^2 \gamma$
are correlated. The correlation in ($\sin 2\alpha$ - $\sin 2\beta$)
is shown in Fig.~\ref{alphabeta2}. Finally, in the SM the
relation $\alpha+\beta+\gamma=\pi$ is satisfied. However, note that the
allowed range for $\beta$ is rather small. Thus,
there is a strong correlation between $\alpha$ and $\gamma$ \cite{AL96}.
  
 It is seen from this figure that 
the smallest value of $\sin 2\beta$ occurs in a small region
of parameter space around $\sin 2\alpha\simeq 0.8$-0.9. Excluding this
small tail, one expects the integrated 
CP-asymmetry in $\bdbarp\to J/\psi K_S$ to be
at least 20\% (i.e., $\sin 2 \beta > 0.4)$, with the central value
estimated as $A(J/\psi K_s)=(30 \pm 7)\%$ \cite{AL96}.
 Less satisfactory at present
is the prediction for the asymmetry related to $\sin 2\alpha$, for which
practically all values are allowed by the fits, including the one
$\sin 2\alpha =0$. If the preferred solution of nature is in the
vicinity of $\sin (2 \alpha)=0$, it is improbable that the
asymmetry related to this quantity will ever be measured. However,
even if $\sin (2 \alpha)$ is not measured at HERA-B, a measurement of
$\sin (2 \beta)$ and a demonstration that $\sin (2 \alpha) \ll \sin(2 \beta)$
will lead to non-trivial constraints on the unitarity triangle. Such a
scenario will also rule out the so-called superweak theory of CP violation
\cite{superweak}, in which case one has the relation $\sin (2 \alpha)=  
\sin(2 \beta)$.

Returning to the CP-asymmetry in the decay $\bdbarp\to J/\psi K_s$,
we recall that the time dependent asymmetry is given by
$$\frac{n(t) - \overline n(t)}{n(t) + \overline n(t)} =
D \sin{2\beta} \sin{xt} \,,
$$ 
where $n(t)$ and $\overline n(t)$ are the time dependent rates
for the decay of a $B^0$ ($\overline B^0$)
to decay into $J/\psi K_s$. $D$ is a dilution factor
which accounts for imperfect tagging.

The accuracy on $\sin{2\beta}$ is given by
$$
\Delta \sin{2\beta} \propto \frac{1}{P}
\frac{1}{\sqrt{N_{B^0}}}
$$
where $P = D \sqrt{\epsilon_{tag}}$ is the tagging power,
$\epsilon_{tag}$ the efficiency to get a tag of the $B^0$ meson
and $N_{B^0}$ the number of reconstructed $B^0$ mesons.
A potential enlargement in the CP reach of HERA-B could be achieved
by an increase in the proton energy at HERA and thus an increase
in the rate of produced $B^0$ mesons. This scenario,
which would substantially improve the 
$\mbox{Signal}/\mbox{Background}$ ratio for $B$ physics at HERA-B,
is, however, somewhat unlikely. 
 A reduction of the error on $\sin{2\beta}$
could however come from reconstructing other $B^0$ decays which also measure
$\sin{2\beta}$. This was studied at this workshop for the decays 
$ \bdbarp \rightarrow \chi_c K^0$ by Misuk and Belyaev and for
$ \bdbarp \rightarrow J/\psi K^{*0}$ by Barsuk
\cite{misuk1}. These studies showed that one could expect a gain in 
statistics of about $20\%$ including both these decays and the favourite 
mode $B \to J/\psi K_s$. 

An increase in the tagging power $P$
and thus a smaller error on $\sin{2\beta}$ could also 
come from new tagging 
techniques. For this purpose the decay 
$B^{**+}\rightarrow \pi^+ B^{0}$ was studied
by Kagan and Shepherd-Themistocleous in the HERA-B environment
\cite{claire}. This tagging is particularly useful
because it is not spoiled by $B \overline B$ mixing. The analysis gave
promising results with a tagging power of $P=0.21$ compared to
the tagging with primary leptons which yielded $P=0.17$. Misuk
investigated the possibility to tag the flavour of the $B$
mesons by using cascade leptons from the decay chain $B\rightarrow
D\rightarrow \ell^\pm$ and obtained for this tagging method 
a tagging power of $P=0.08$ \cite{misuk2}. 
In summary, these studies undertaken to increase the sensitivity of the 
HERA-B experiment for $\sin{2\beta}$ showed that a gain in the
statistical power of about 30$\%$ is possible.

% This is Figure 3

\subsection{$\delms$ (and $\xs$) and the Unitarity Triangle}

Mixing in the \bsbsbar\ system is quite similar to that in the \bdbdbar\
system in the SM in which the \bsbsbar\ and \bdbdbar\ 
box diagrams are dominated by $t$-quark exchange.
Using the fact that $\vert V_{cb}\vert=\vert V_{ts}\vert$ (Eq.~\ref{CKM}),
it is clear that one of the sides of the unitarity triangle, $\vert
V_{td}/\lambda V_{cb}\vert$, can be obtained from the ratio of $\delmd$ and
$\delms$,
\beq
\frac{\delms}{\delmd} =
 \frac{\hat{\eta}_{B_s}M_{B_s}\left(\fbbs\right)}
{\hat{\eta}_{B_d}M_{B_d}\left(\fbb\right)}
\left\vert \frac{V_{ts}}{V_{td}} \right\vert^2.
\label{xratio}
\eeq
Here $\hat{\eta}_{B_s}=\hat{\eta}_{B_d} =0.55$ is the perturbative QCD
correction factor  
\cite{Burasetal}.
 The only real uncertainty (apart from the CKM matrix element ratio which
we would like to determine) 
is the ratio of hadronic matrix elements $\fbbs/\fbb$.
 Using the determination of $A$ given
earlier, $\tau(B_s)= 1.52 \pm 0.07 ~ps$,  and $\overline{\mt}=165 \pm 9$ 
GeV, one gets
\begin{eqnarray}
\delms &=& \left(12.8 \pm 2.1\right)\frac{\fbbs}{(230~\mbox{MeV})^2} 
~(ps)^{-1}~, \nonumber \\
\xs &=& \left(19.5 \pm 3.3\right)\frac{\fbbs}{(230~\mbox{MeV})^2}~.
\end{eqnarray}
The choice $f_{B_s}\sqrt{\hat{B}_{B_s}}= 230$ MeV corresponds to the
central value given by the lattice-QCD estimates, and with this the fits
in \cite{AL96} 
give $\xs \simeq 20$ as the preferred value in the SM. Allowing the
coefficient to vary by $\pm 2\sigma$, and taking the central value for
$f_{B_s}\sqrt{\hat{B}_{B_s}}$, this gives 
\begin{eqnarray} 
12.9 &\leq & \xs \leq 26.1~, \nonumber\\
8.6 ~(ps)^{-1} &\leq & \delms \leq 17.0 ~(ps)^{-1}~.
 \label{bestxs}
\end{eqnarray}
It is difficult to ascribe a confidence level to this range due to the
dependence on the unknown coupling constant factor. All one can say is that
the standard model predicts large values for $\delms$ (and hence $\xs$).
The present experimental limit from the combined fit of the ALEPH 
\cite{ALEPHxs} and OPAL experiments $\delms > 9.2
~(ps)^{-1}$ \cite{Gibbons96} is better than the lower bound
on this quantity obtained from the CKM fits given above.
 In particular, the LEP-bound, expressed as $\delms/\delmd > 19.0$, 
removes the (otherwise allowed)
large-negative-$\rho$ region, leaving the reduced parameter space $(-0.25 
\leq \rho \leq 0.35$, $0.25 \leq \eta \leq 0.52)$ as the presently
allowed one \cite{AL96}. In terms of the ratio $\absvtd/\absvts$,
this implies $\absvtd/\absvts < 0.29$, to be compared with the
central value from the CKM fits $\absvtd/\absvts =0.24$ \cite{AL96}.
 The constraints on the unitarity triangle
from $\delms$ will become more pronounced with improved data.  
With the present HERA-B detector,
one expects to reach a sensitivity $\xs \simeq 17$ (or
$\delms \simeq 11~ps^{-1})$ combining various $B_s$ reconstruction 
and tagging
techniques and data for three years \cite{JThom96}. This overlaps
with the $x_s$-range expected in the SM, though
it is somewhat on the lower side. To completely cover the estimated 
$x_s$-range in eq.~(\ref{bestxs}), one must strive to increase the 
experimental sensitivity to $x_s=26$.

%%%%%%%%%%%%%%%%%%%%%%%%%%%%%%%%
\section{Rare $B$ Decays in the SM and at HERA-B}

The FCNC $B$ decay with the largest branching ratio in the SM,
 $\BGAMAXS$, has been observed by the CLEO
collaboration \cite{CLEOrare2} through the measurement of the photon 
energy spectrum in the high $E_\gamma$-region.
 This was preceded by the observation of the exclusive 
decay $\BGAMAKSTAR$ by the same collaboration \cite{CLEOrare1}. The present 
measurements give \cite{CLEOwarsaw}:
${\cal B}(\BGAMAXS) = (2.32\pm 0.57\pm 0.35)\times 10^{-4}$ and
${\cal B}(\BGAMAKSTAR) = (4.2\pm 0.8 \pm 0.6)\times 10^{-5}$,
yielding an exclusive-to-inclusive ratio:
$R_{K^*} = \Gamma(\BGAMAKSTAR)/\Gamma(\BGAMAXS)=(18.1\pm 6.8)\% $.
In the SM the decay rates determine the ratio of the CKM matrix 
elements $\Vtsabs/\Vcbabs$ and the quantity $R_{K^*}$ provides information
on the decay form factor in $\BGAMAKSTAR$.
It is important to undertake independent measurements of the
above-mentioned decays and related processes elsewhere. 

A Monte Carlo study by  Saadi-L\"udemann described 
in these proceedings \cite{Saadi96} shows that the large-$p_T$ photons
emerging from the decays $B \to K^* + \gamma$ can, in principle,
be distinguished from the background events at HERA-B,
which range out earlier in $p_T$. Only slight modifications
to the present HERA-B trigger scheme are necessary. This argument also
holds for the photons from  the inclusive decay $\BGAMAXS$, as
the photon $p_T$-spectrum is rather hard and the additional
requirement of a large-$p_T$ charged track accompanying the energetic  
photon will be fulfilled by these decays. 
Since  $b$-quark fragments include typically $20\%$ of the time a $B_s^{0}$
meson or a $b$ baryon (henceforth generically called $\Lambda_b$), the
FCNC rare decays of the $B_s^{0}$ meson and $\Lambda_b$ baryon
will be new and valuable additions to this field which cannot
be studied at $e^+e^-$ threshold machines, optimized to operate
at the $\Upsilon(4S)$ resonance. The 
approximate equality of the inclusive radiative
branching ratios for the decays $B^\pm \to X_{s}^\pm + \gamma,
 ~B_d^{0} \to X_{s}^{0} + \gamma, ~B_s^{0} \to X_{(s\bar{s})}^{0} + 
\gamma$ and $\Lambda_b \to (\Lambda+X) + \gamma$ will test the hypothesis 
that these decays are indeed dominated by short-distance physics.
Given conducive triggers, HERA-B has the potential of contributing
significantly to the field of rare $B$ decays. Here, we
summarize some  representative examples which can be studied
at HERA-B, given the present and planned triggers and assuming that
$10^{9}$ $b\bar{b}$ pairs will be produced in three years of data taking
at HERA-B.
   
\subsection{Inclusive decay rates \bgamaxs ~and \bgamaxd}
The leading contribution to $b \to s +\gamma$ arises
at one-loop from the so-called penguin diagrams. With the help of the
unitarity of the CKM matrix,
the decay matrix element in the lowest order can be written as:
\begin{equation} \label{e2}
 {\cal M }(b \to s ~+\gamma)
    = \frac{G_F}{\sqrt{2}}\,\frac{e}{2 \pi^2} \,\lambda_{t}
   \,(F_2 (x_t)-F_2(x_c))\, q^\mu \epsilon^\nu \bar{s} \sigma_{\mu \nu}
      (m_bR ~+ ~m_sL)b ~.
 \end{equation}
where $x_i= ~m_i^2/m_W^2$, and
$q_\mu$  and $\epsilon_\mu$ are, respectively, the photon four-momentum
and polarization vector, and $\lambda_t=V_{tb} V_{ts}^*$.
The (modified) Inami-Lim function $F_2(x_i)$ derived from the (1-loop) 
penguin diagrams \cite{InamiLim} can be seen in \cite{ALI96}.
 The measurement of the branching ratio for $\BGAMAXS$ can be 
readily interpreted in terms of the CKM-matrix element product
$\lambda_t/\Vcbabs$ or equivalently $\Vtsabs/\Vcbabs$.
For a quantitative determination
of $\Vtsabs/\Vcbabs$, however,  QCD radiative
corrections have to be included and
the contribution  of the so-called long-distance effects estimated.
This has been reviewed in \cite{ALI96}, yielding:
\begin{equation}\label{smbsgbrf}
{\cal B} (\BGAMAXS )= (3.20 \pm 0.58) \times 10^{-4},
\end{equation}
which is compatible with the present measurement
${\cal B} (\BGAMAXS )= (2.32 \pm 0.67) \times 10^{-4}$ \cite{CLEOrare2}.
Expressed in terms of the CKM matrix element ratio, one gets \cite{ALI96}
\begin{equation}\label{vtscb}
\frac{\Vtsabs}{\Vcbabs} = 0.85 \pm 0.12 (\mbox{expt}) \pm 0.10 (\mbox{th}),
\end{equation}
which is within errors consistent with unity, as expected from the
unitarity of the CKM matrix.

 Since the masses and lifetimes of the $B^\pm, ~B_d^{0}$, and $B_s^{0}$
mesons are very similar, the branching ratio quoted above holds (within
minor differences) for all three $B$ mesons. The branching ratio for the
$\Lambda_b$-baryon will be reduced by the ratio of the lifetimes. One 
estimates,
 \begin{eqnarray}
{\cal B}(\Lambda_b\to (\Lambda+X)\gamma) &=&
{\cal B}(B_{d}\to X_{s}\gamma)
\left[\frac{\tau(B_{d})}{\tau(\Lambda_b)}\right]\nonumber\\
&=& (2.5 \pm 0.6) \times 10^{-4} ~,
\end{eqnarray}
where we have used $\tau(B_{d})/\tau(\Lambda_b)=0.78 \pm 0.04$ 
\cite{Richman96}.

\par
The theoretical interest in studying the 
(CKM-suppressed) inclusive radiative decays
\bgamaxd\ lies in the first place in the 
 possibility of determining the parameters of the CKM
matrix.
 With that goal in view, one of the relevant quantities in the
decays $B \to X_d + \gamma$ is the end-point photon energy spectrum
which has to be measured requiring that
 the hadronic system $X_d$ recoiling against the
photon does not contain strange hadrons, so as to suppress the
 large-$E_\g$ photons from the decay $\BGAMAXS$. This requires, in 
particular, a good $K/\pi$-separation.
 Assuming that this is feasible,
one can determine  from the ratio of the decay rates
$\BBGAMAXD/\BBGAMAXS$ the CKM-Wolfenstein parameters $\rho$ and $\eta$.
 To get an estimate of the inclusive branching ratio at present,
the CKM parameters $\rho$ and $\eta$ have to be constrained from the
unitarity fits discussed above.
 Taking the preferred
values of the fitted CKM parameters from eq.~(\ref{bestrhoeta}), one gets 
\cite{ag2,aag96}   \begin{equation}
 \BBGAMAXD = 1.63 \times 10^{-5},
\end{equation}
whereas $\BBGAMAXD =8.0 \times 10^{-6}$ and $2.8 \times 10^{-5}$ for the 
other 
two extremes $\rho=0.35, ~\eta=0.50$ and $\rho=-\eta=-0.25$, respectively.
Therefore, one expects $O(10^4)$ $\BGAMAXD$ events at HERA-B, which taking
into account an estimated trigger and reconstruction
 efficiency of $1\%$ would yield $O(10^2)$ reconstructed $B$ decays
 of this kind. However, one will have  to suppress the background
from the dominant $\BGAMAXS$ decays which requires further study.
\subsection{${\cal B}(B \to V + \gamma )$ and constraints on the CKM 
parameters}

Exclusive radiative
 $B$ decays $B \to V + \gamma$, with $V=K^*,\rho,\omega$, are also 
potentially
very interesting from the point of view of determining the CKM parameters
\cite{abs93}. The extraction of these parameters would, however,  involve a 
trustworthy 
estimate of the SD- and LD-contributions in the decay amplitudes.

  The SD-contribution in the 
 exclusive decays $(B^\pm, B^{0}) \to (K^{*\pm}, K^{* 0})+ \gamma$,
$(B^\pm, B^{0}) \to (\rho^\pm,\rho^{0}) + \gamma$,
$B^{0} \to \omega + \gamma$  and the
corresponding $B_s$ decays, $B_s \to \phi + \gamma $, and
$B_s \to K^{* 0} + \gamma $,
involve the magnetic moment operators \cite{ALI96}.
The transition form factors governing these decays
 can be generically  defined as:
\be
 \langle V,\lambda |\frac{1}{2} \bar \psi \sigma_{\mu\nu} q^\nu b
 |B\rangle  =
     i \epsilon_{\mu\nu\rho\sigma} e^{(\lambda)}_\nu p^\rho_B p^\sigma_V
F_S^{B\rightarrow V}(0).
\label{defF}
\ee
Here $V$ is a vector meson
with the polarization vector $e^{(\lambda)}$,
$V=\rho, \omega, K^*$ or $\phi$;
$B$ is a generic
$B$-meson $B^\pm, B^{0}$ or $B_s$, and $\psi$ stands for the
field of a light $u,d$ or $s$ quark. The vectors $p_B$, $p_V$ and
$q=p_B-p_V$ correspond to the 4-momenta of the initial $B$-meson and the
outgoing vector meson and photon, respectively. Keeping only the 
SD-contribution leads to obvious relations 
among the exclusive decay rates, 
\be
\frac{\Gamma ((B^\pm,B^{0}) \to (\rho^\pm,\rho^{0}) + \gamma)}
     {\Gamma ((B^\pm,B^{0}) \to (K^{*\pm},K^{* 0}) + \gamma)} 
  \simeq \kappa_{u,d}\left[\frac{\Vtdabs}{\Vtsabs}\right]^2 \,,
\label{SMKR}
\ee
where
 $\kappa_{i} \equiv [F_S(B_i \to \rho \gamma)/F_S(B_i \to K^* 
\gamma)]^2 \Phi_{u,d}$ and
 $\Phi_{u,d}$ is a phase-space factor which in all cases is close to 1.
Likewise, using the SD-contribution and isospin symmetry yields
\beq\label{ratio2}
\Gamma(B^\pm \to \rho^\pm \gamma)=2 ~\Gamma(B^{0}\to \rho^0  \gamma)
    = 2 ~\Gamma (B^{0} \to \omega  \gamma)~.
\label{isospin}
\eeq

 If the SD-amplitudes were the only contributions, the measurements of the
 CKM-suppressed radiative decays $(B^\pm,B^0) \to (\rho^\pm, \rho^0) + 
\gamma , ~B^0 \to \omega + \gamma$ and $B_s^0 \to K^{*0} + \gamma$ could be
used in conjunction with the decays $(B^\pm,B^0) \to (K^{*\pm},K^{*0}) + 
\gamma$ to determine one of the sides of the unitarity triangle.
The present experimental upper limits on the CKM ratio
$\Vtdabs/\Vtsabs$ from radiative $B$ decays 
are indeed based on this assumption. The present limits on some of the
 decay modes are reviewed in \cite{ALI96}, which yield at 90\% 
C.L.\cite{CLEOwarsaw}: %
\be
\left\vert {V_{td} \over V_{ts}} \right\vert \leq 0.45 - 0.56~,
\ee
depending on the models used for the $SU(3)$ breaking effects
in the form factors. The estimated range for this ratio is
$0.15 \leq \Vtdabs/\Vtsabs \leq 0.29$, which implies that an
improvement of a factor of 3 - 10 in the experimental sensitivity
would result in measurements of several CKM-suppressed radiative decay 
modes. 

  The possibility of significant
LD-contributions in
radiative $B$ decays from the light quark intermediate states
has been raised in a number of papers.
The LD-contributions in $B \to V + \gamma$  are
induced by the matrix elements of the
four-Fermion operators (see \cite{ALI96} for definitions and references).
Their amplitudes necessarily involve other CKM matrix elements and hence the
simple factorization of the decay rates in terms of the CKM factors
involving $\Vtdabs$ and $\Vtsabs$ no longer holds thereby
invalidating the relations (\ref{SMKR}) and (\ref{isospin}) given above.
The modified relations have been worked out in  \cite{wyler95,ab95}.
Combining the estimates for the LD- and SD-form factors in
\cite{ab95} and
\cite{abs93}, respectively, and restricting the Wolfenstein
parameters in the range $-0.25 \leq \rho \leq 0.35$ and $ 0.2 \leq \eta
\leq 0.5$, as discussed above, the
following estimates for the absolute branching ratios have been
given in  \cite{ALI96}:
\begin{eqnarray}\label{ratio4}
{\cal B}(B^\pm\to \rho^\pm\gamma)
&=& (1.5 \pm 1.1) \times 10^{-6} ~,
\nonumber\\
{\cal B}(B^{0}\to \rho\gamma) &\simeq& {\cal B}(B^{0}\to \omega \gamma)
= (0.65 \pm 0.35) \times 10^{-6} ~,
\end{eqnarray}
where we have used the experimental value for the branching ratio
${\cal B} (B \to K^* + \gamma)$
\cite{CLEOrare1},
adding the errors in quadrature. The large error reflects the poor
knowledge of the CKM matrix elements and hence experimental determination
of these branching ratios will put rather stringent constraints on the
Wolfenstein parameters, in particular $\rho$.

In addition to studying the radiative penguin decays of the $B^\pm$ and
$B^0$ mesons discussed above, HERA-B will be in a 
position to study the
corresponding decays of the $B_s^0$ meson and $\Lambda_b$ baryon, such as
$B_s^0 \to \phi + \gamma$ and $\Lambda_b \to \Lambda + \gamma$, which have
not been measured so far. We list below the branching ratios in a number of
interesting decay modes calculated in the QCD sum rule approach in 
\cite{abs93}.
\begin{eqnarray}\label{ratio6}
{\cal B}(B_s\to \phi\gamma)
&\simeq& {\cal B}(B_d\to K^* \gamma)
= (4.2 \pm 2.0) \times 10^{-5} ~,
\nonumber\\
\frac{{\cal B}(B_s\to K^*\gamma)}{{\cal B}(B_d\to K^*\gamma)}
 &\simeq & (0.36 \pm 0.14)  \frac{\Vtdabs^2}{\Vtsabs^2} \nonumber\\
&\Longrightarrow & {\cal B}(B_s\to K^*\gamma)= (0.75 \pm 0.5) \times 
10^{-6} ~.
\end{eqnarray}

   The branching ratio for the radiative decay $\Lambda_b \to \Lambda 
\gamma$ can be calculated in terms of the
exclusive-to-inclusive function $R_{\Lambda} \equiv \Gamma (\Lambda_b
\to \Lambda + \gamma) /\Gamma(\Lambda_b \to (\Lambda +X) + \gamma)$, 
analogous to the ratio $R_{K^*}$, defined for the $B$ meson decays earlier. 
This would then give:
\begin{equation}
{\cal B}(\Lambda_b\to \Lambda\gamma) = R_{\Lambda}
{\cal B}(\Lambda_b\to (\Lambda +X)\gamma)
= (3.8 \pm 1.5) \times 10^{-5} ~,
\end{equation}
where we have used the SM estimate ${\cal B}(B_{d}\to 
X_{s}\gamma)=(3.2\pm 0.6)\times 10^{-4}$ \cite{ALI96}, assumed 
$R_{\Lambda}=R_{K^*}$, with $R_{K^*} =0.15\pm 0.05$ \cite{abs93} and
$\tau(B_{d})/\tau(\Lambda_b)=0.78 \pm 0.04$ \cite{Richman96}. We note that
a much smaller branching ratio has been calculated in \cite{singer96}.
The CKM-suppressed decay $\Lambda_b^{0} \to n + \gamma$ is related to
the decay $\Lambda_b^{0} \to \Lambda + \gamma$ by the CKM ratio 
$\Vtdabs/\Vtsabs^2$ in the $SU(3)$ limit. Using the central value
$\Vtdabs/\Vtsabs^2= 0.058$, this gives
${\cal B}(\Lambda_b\to n + \gamma)= (2.2\pm 1.0) \times 10^{-6}$.
Estimated counting rates
 for several of the exclusive decay modes in the HERA-B
experimental environment are given in these proceedings \cite{Saadi96},
taking into account the trigger and reconstruction efficiency.
They range between $O(10^2)$ for the CKM-allowed decays and $O(10)$ for
the CKM-suppressed decays.  

\subsection{Inclusive rare decays $B \to X_s \ell^+ \ell^-$ in the SM}

The decays \bxsll, with $\ell=e,\mu,\tau$, provide a more sensitive search
strategy for finding new physics in rare $B$ decays
than for example the decay \bxsg , which constrains
the magnitude of the effective Wilson coefficient of 
the magnetic moment operator, $C_7^{\mathit{eff}}$.
The sign of $C_7^{\mathit{eff}}$ (which is negative in the SM but in general
depends on the underlying physics) is not
determined by the measurement of ${\cal B}(\BGAMAXS)$ alone.
It is known (see for example \cite{AGM94}) that
in supersymmetric (SUSY) models, both the negative and positive signs are 
allowed as one scans over the allowed SUSY parameter space.
We recall that the \bxsll ~amplitude in the standard model (as well as in 
SUSY and multi-Higgs models)
has two additional terms, arising from the two FCNC four-Fermi operators.
Calling their coefficients $C_{9}$ and $C_{10}$, it has been argued in
\cite{AGM94} that the signs and
magnitudes of all three coefficients $C_7^{\mathit{eff}}$, $C_{9}$ and 
$C_{10}$
can, in principle,  be determined from the decays $\BGAMAXS$ and \bxsll .

\begin{figure}[htb]
\vskip -0.1truein
\centerline{\psfig{figure=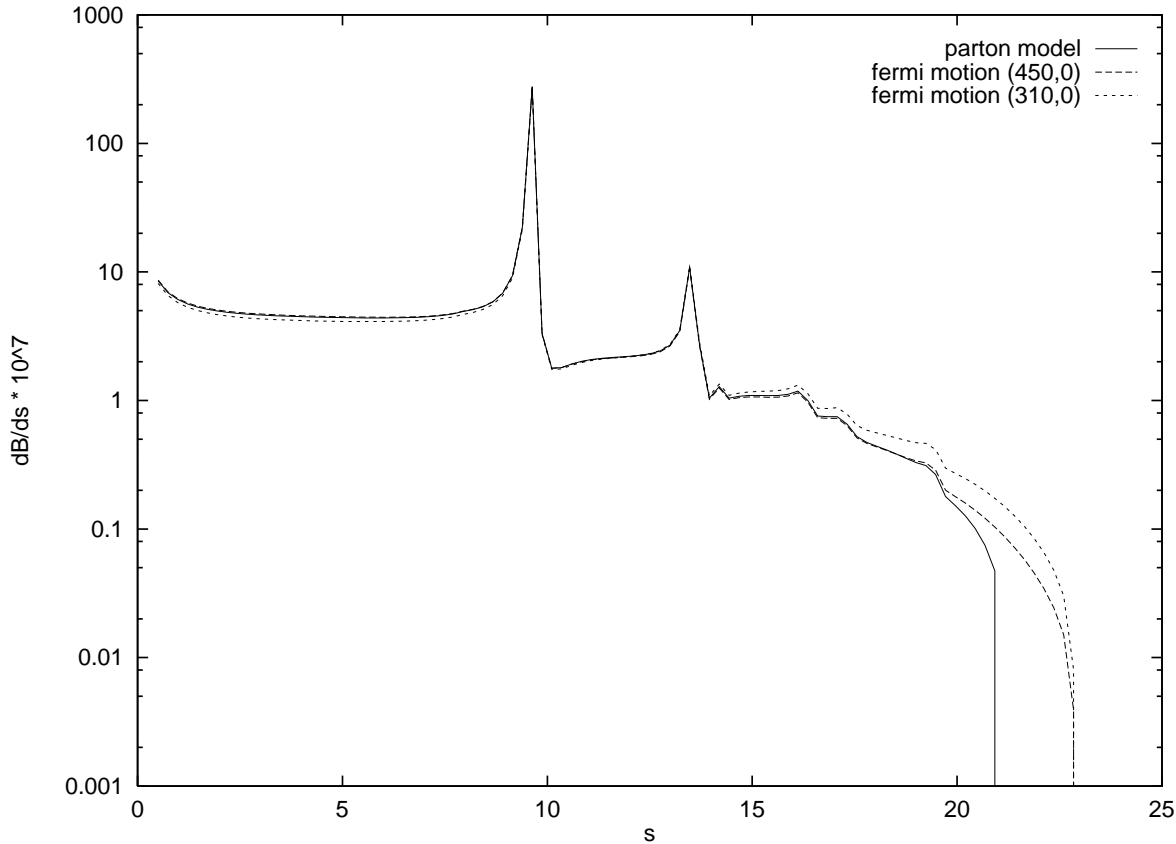,height=16.0cm,angle=270}}
\vskip -0.1truein
\caption[]{
Dilepton invariant mass distribution in $B \to X_s \ell^+ \ell^-$ in
 the SM including
next-to-leading order QCD correction and LD effects. The solid curve
corresponds to the parton model and the short-dashed and long-dashed
curves correspond to including the Fermi motion effects. The
values of the Fermi motion model are indicated in the figure.}
\label{fig:lddb}
\end{figure}

The differential decay rate in the dilepton invariant mass in \bxsll 
can be expressed in terms of the semileptonic branching ratio
${\cal B}_{sl}$, 
 \begin{eqnarray}
	{{\rm d}{\cal B}(\hat{s}) \over {\rm d}\hat{s}} & = &
		{\cal B}_{sl} \frac{\alpha^2}{4 \pi^2} \frac{ 
		\lambda_t^2}{\Vcbabs^2} \frac{1}{f(\hat{m}_c) \kappa(\hat{m}_c)}
		u (\hat{s}) \left[ \vphantom{\frac{1}{1}}
		\left( |C_9^{\mathit{eff}}(\hat{s})|^2
 		+ C_{10}^2 \right) \alpha_1 (\hat{s},\hat{m}_s)
		\right. \nonumber \\
& & \left. + \frac{4}{\hat{s}} (C^{eff}_7)^2 \alpha_2 (\hat{s},\hat{m}_s)
+ 12 \alpha_3 (\hat{s},\hat{m}_s) C^{eff}_7 {\cal 
\Re}(C_9^{\mathit{eff}}(\hat{s}))
		\right] ,
	\label{eqn:dbrs}
\end{eqnarray}
with  
$\hat{s}=s/m_b^2$, $u(\hat{s})=\sqrt{\left[\hat{s}-(1+\hat{m_s})^2\right]
\left[\hat{s}-(1-\hat{m_s})^2 \right]}, ~\hat{m}_i=m_i^2/m_b^2$, and
the functions $f(\hat{m}_c), ~ 
~\kappa(\hat{m}_c)$, and $\alpha_i$ can be seen in \cite{ALI96}. 
Here ${\cal \Re}(C_7^{\mathit{eff}})$ represents the real part of 
$C_7^{\mathit{eff}}$.
A useful quantity is the  differential FB asymmetry in the c.m.s. of the
dilepton
defined in refs. \cite{amm91}:
\begin{equation}\label{FBasym}
\frac{d {\cal A}(\hat{s})}{d\hat{s}} = \int_0^1 \frac{d{\cal B}}{dz}
                                      -\int_0^{-1} \frac{d{\cal B}}{dz},
\end{equation}
where $z=\cos \theta$, which can be expressed as:
\begin{eqnarray}
	{{\rm d}{\cal A}(\hat{s}) \over {\rm d}\hat{s}} & = &
		- {\cal B}_{sl} \frac{3 \alpha^2}{4 \pi^2}
                \frac{1}{f(\hat{m}_c)} u^2 (\hat{s})
		C_{10} \left[ \hat{s}{\cal \Re} ( C_9^{\mathit{eff}}(\hat{s})) +
		2 C^{eff}_7 (1 + \hat{m}_s^2) \right] .
	\label{eqn:dasym}
\end{eqnarray}
 The Wilson coefficients
$C^{eff}_7$, $C^{eff}_9$ and $C_{10}$ appearing in the above equations
can be determined from data by solving the partial branching ratio
${\cal B}(\Delta \hat{s})$ and partial FB asymmetry
${\cal A}(\Delta \hat{s})$, where $\Delta \hat{s}$ defines an
interval in the dilepton invariant mass \cite{AGM94}.
From the experimental point of view, the normalized FB-asymmetry, which is
defined as
\begin{equation}\label{NFBasym}
\frac{d A(\hat{s})}{d\hat{s}} = \frac{\int_0^1 \frac{d{\cal B}}{dz}
                                      -\int_0^{-1} \frac{d{\cal B}}{dz}}
                                     {\int_0^1 \frac{d{\cal B}}{dz}
                                      +\int_0^{-1} \frac{d{\cal B}}{dz}}\,,
\end{equation}
is a more useful quantity.
 Following branching ratios for the SD-piece have been estimated in
\cite{AHHM96}:
 \begin{eqnarray}\label{brbsll}
{\cal B}(\bxsee) &=& (8.4 \pm 2.2) \times 10^{-6}, \nonumber\\
{\cal B}(\bxsmm) &=& (5.7 \pm 1.2) \times 10^{-6}, \nonumber\\
{\cal B}(\bxstt) &=& (2.6 \pm 0.5) \times 10^{-7}. 
\end{eqnarray}
The present best upper limit is ${\cal B}(\bxsmm) < 3.6 \times 10^{-5}$
at $(90\%$ C.L.) by the D0 collaboration \cite{D0warsaw}, and there are
no useful limits on the other two decay modes.
 To get the physical decay
rates and distributions, one has to include the LD-contributions
(which are dominated by the $J/\psi$ and $\psi^\prime$ resonances) and
the hadronic wave function effects. These aspects have been recently studied
in \cite{AHHM96}, using the Fermi motion model \cite{AlipietAlt} which 
depends
on two parameters $p_F$ (the $b$-quark kinetic energy) and $m_q$ (the
spectator quark mass in $B=\bar{b}q$).
We show here the resulting invariant dilepton mass spectrum
in Fig.~\ref{fig:lddb}, from  which it is obvious that   
only in the dilepton mass region far away from
the resonances is there a hope of extracting the Wilson coefficients
governing the short-distance physics. The region below the $J/\psi$ resonance
is well suited for that purpose for HERA-B as the dilepton invariant 
mass distribution here is dominated by the SD-piece. 
% 
%
% This is Fig. 4.
The prerequisite for measurements at HERA-B is a lower dilepton mass trigger
than is the case now which is optimized at the $J/\psi$ mass, 
otherwise the SD-piece in $\bxsmm$ will be harder to extract.
Including the LD-contributions, following branching ratio has been
estimated for the dilepton mass range $2.10$ GeV$^2 ~\leq s \leq 2.90$ 
GeV$^2$ in \cite{AHHM96}:
\be \label{brbslld}
{\cal B}(\bxsmm) = (1.3 \pm 0.3) \times 10^{-6}, 
\ee
with ${\cal B}(\bxsee) \simeq {\cal B}(\bxsmm)$. The 
normalized FB-asymmetry is estimated to be 
in the range $10\%$ - $27\%$.
These branching ratios and the FB asymmetry
are expected to be measured within the next several years at
HERA-B with few tens of events \cite{Saadi96} 
and other forthcoming $B$ facilities.
 In the high invariant mass region, the short-distance contribution 
dominates. However, the rates are down by roughly an order of magnitude
compared to the region below the $J/\psi$-mass. Estimates of the
branching ratios are of $O(10^{-7})$, which should be accessible at
the LHC.

In conclusion, the semileptonic FCNC decays $B \to X_s \ell^+ \ell^-$
(and the related exclusive decays) will provide very precise tests
of the SM in flavour physics. They may also reveal new physics beyond
the SM. The MSSM model is a good case study where measurable deviations
from the SM are anticipated are anticipated and worked out
 \cite{AGM94,CMW96}.

\subsection{ Other topics in $B$ physics}

With large samples of $B$ hadrons available, also other precision 
measurements involving $B$ decays are feasible at HERA-B, e.g.
precise determination of $\absvcb$, $B$-hadron lifetimes 
and the $B_d^0$ - $\overline{B_d^0}$ mixing ratio $x_d$.
 Present measurements of $\absvcb$ are at $\pm 7\%$ \cite{Gibbons96}, and the
lifetimes of the specific $B$ hadrons are 
(all in $ps$):
$\tau(B^\pm)=1.65 \pm 0.04, ~\tau(B_d^0)=1.55 \pm 0.04, ~\tau(B_s^0)=1.52
\pm 0.07$ and $\tau(\Lambda_b)=1.21 \pm 0.06$ \cite{Richman96}.
Theoretical estimates based on the QCD-improved parton model predict
almost equal lifetimes of the charged $(B^\pm)$
 and neutral $(B_d^0,B_s^0)$ mesons and the $\Lambda_b$
baryons. Power corrections will split these lifetimes but only moderately.
There exists a mild embarrassment for present theoretical estimates in 
that the ratio
$\tau(\Lambda_b)/\tau(B_d) = 0.78 \pm 0.04$ is significantly lower than 
unity (expected to be $> 0.9$ in most estimates).
To check these models, lifetime have to be measured very accurately,
in particular of the $\Lambda_b$ and $B_s^0$. Of some theoretical interest
is also the lifetime difference between the two mass eigenstates $\Delta 
\Gamma (B_{s,1}-B_{s,2})$, which is expected to be $O(15\%)$ \cite{Beneke96}.
Kreuzer showed that lifetime measurements at one to two percent level are 
feasible for the $B^\pm$ and the neutral $B^0$ mesons, and at
$\pm 5\%$ for $\Lambda_b$ and $B_s^0$ at HERA-B \cite{kreuzer}.
Another interest lies in the precise determination of $\Delta m_d$,
\cite{kreuzer}. Although the combined
LEP and ARGUS/CLEO measurements have reached an accuracy of a few percent
with $\delmd = 0.464 \pm 0.018 ~(ps)^{-1}$ \cite{Gibbons96},
HERA-B would be able to contribute to these measurement with comparable
errors and very different systematic effects, as it was demonstrated
by Kreuzer in this workshop.  

  The question of producing and detecting the mesons $B_c\equiv\bar{b}c$
(and its charge conjugate) at
HERA-B and HERA$(ep)$ was discussed by Baranov, Ivarsson, Mannel, and 
R\"uckl at this workshop. This is an interesting object to study,
as both the $\bar{b}$ and $c$ quarks can decay independently and
$O(5\%)$ decays would take place via the annihilation diagram. The decay 
products involve final states such as $J/\psi +(\pi,\rho,A_1,...)$ and the
semileptonic decays such as $J\psi \ell^+ \nu_\ell$, which are measurable
at HERA-B and HERA$(ep)$. Unfortunately, the production cross sections are
small at HERA in both the $ep$ and $pp$ modes. Typical estimates are:
$\sigma(pp \to B_c \bar{c} b X) \simeq 10 ~fb$ at the HERA-B energy
$\sqrt{s}\simeq 40$ GeV, with the cross section in the vector meson 
$(B_c^*)$ mode a factor 2 - 3 larger. This, for example for the $J/\psi 
\pi$ mode yields $\sigma (B_cX) \cdot {\cal B}(J/\psi \pi) = O(10^{-2}) 
fb$, putting its detection beyond the integrated HERA-B luminosity. At
HERA$(ep)$, the production cross section is estimated as     
$\sigma(ep \to B_c \bar{c} b X) \simeq 1 ~pb$, making it well nigh
impossible to detect the $B_c$-meson
even with an integrated luminosity of $250 ~(pb)^{-1}$.

\section{Rare Decays, $D^0$ - $\overline{D^0}$ Mixing and CP Violation}

  As discussed by Eichler and Frixione in these proceedings, the measured
charm hadron photoproduction cross section at HERA is close to one microbarn.
At present, a total efficiency of $10^{-4}$ of charmed hadron reconstruction 
via $D^*$-tagging has been achieved at HERA, which is expected to go up 
to $O(10^{-3})$ by adding various useful decay modes of $D^0$ and having
the benefit of a vertex
detector. With an integrated luminosity of $250 ~(pb)^{-1}$,
and including the $D^\pm$ mesons, this could
yield up to $10^6$ reconstructed $D^0 (\overline{D^0})$ and$ D^\pm$ 
events with a $S/N \geq 1$. At 
HERA-B, the charmed hadron production cross section is estimated as
 $O(10 ~\mu b)/\mbox{Nucleon}$, consistent with the fixed target
experiments \cite{Appel94}, leading to $O(10^{12})$ charmed hadrons
produced in three years of data taking.
 No detailed study of the charmed hadron 
reconstruction efficiency has been undertaken at HERA-B. Hence, it is
difficult to be quantitative. However, the method of $D^*$-tagging 
coupled with 
vertex resolution studied in the context of HERA will be useful at HERA-B
as well. With an (assumed) overall reconstruction efficiency of 
$O(10^{-5})$ at HERA-B, 
this would lead to $O(10^{7})$ reconstructed charmed events.
  As reviewed by Jeff Appel during this workshop, fixed target experiments
(in particular E791 and E687) have already reconstructed in excess of
$10^5$ charmed hadrons. A programme to reach $O(10^6)$ 
charmed hadrons in fixed target experiments in USA is already in place. 
There are
enticing proposals to get to $O(10^8)$ (or even $10^9)$ charmed hadrons,
though the time scale for beyond $10^6$ 
is difficult to predict. It is clear that both HERA and
HERA-B have formidable tasks ahead in matching the current and
planned performances in the charmed hadron sector.

 The current interest in the charm sector lies in doing what has become
to be known as ``the high impact physics". This means searches for 
rare decays, $D^0$ - $\overline{D^0}$ mixing and CP violation. Some of
the rare $D$ decays were studied during the previous HERA workshop 
\cite{Egli91}. An updated study of $D^0 \to \mu^+ \mu^-$ has been undertaken 
by Grab during this workshop \cite{Grab96}, with the conclusions that 
with an integrated luminosity of $250 ~(pb)^{-1}$ at HERA, a sensitivity of
$2.5 \times 10^{-6}$ will be reached in this channel. The present upper
limit on this decay mode is  $7.6 \times 10^{-6}$ \cite{PDG96}.
This implies a factor of 3 improved reach at HERA. The sensitivities
in the other leptonic modes $D^0 \to e^+ e^-, ~D^0 \to \mu^\pm e^\mp$ and
the semileptonic decays $D^0 \to \rho^0 e^+ e^-, ~\rho^0 \mu^+ \mu^-$, as
well as the analogous charged decays $D^\pm \to \pi^\pm e^+ e^-, 
~D^\pm \to \pi^\pm \mu^+ \mu^-$
were studied during the previous HERA workshop and can be seen there.
In the meanwhile, upper limits on several of these decay modes have moved
up significantly \cite{PDG96}, leaving a reduced window for searches at
HERA.

  Finally, we note that both mixing and CP violation in the charm sector
are too small to be measured, 
 if the SM is the only source of such transitions. Typically
in the SM, one has the following scenario \cite{Georgi}:
\begin{equation}
\frac{\Delta M_D}{\Gamma_D} \simeq O(10^{-5}),
 ~~\frac{\Delta \Gamma_D}{\Delta M_D} \ll 1,
\end{equation}
 with the CP-violating quantity ${\it Im}(\Delta M_d/\Gamma_D)$ negligible.
The feature $\Delta \Gamma_D/\Delta M_D \ll 1$ will hold in all extensions
of the SM, as the decay rates are determined essentially by the tree 
diagrams and one does not anticipate large enhancements here.
However, in a number of extensions of the SM, the quantity $\Delta 
M_D/\Gamma_D$ may receive additional contributions pushing it
close to its present upper limit \cite{Hewett}. In 
addition, in
some theoretical scenarios, one has ${\it Im}(\Delta M_d/\Gamma_D) \simeq
{\it Re}(\Delta M_d/\Gamma_D)$, implying also measurable CP-violation. This 
could manifest itself in the differences in the time-dependent and
time-integrated rates, leading to CP violating asymmetries in $D^0 (t) 
\to f$ and $\overline{D^0} (t) \to
\bar{f}$, where $f$ and $\bar{f}$ are CP-conjugate states. Examples of
such extensions are: SUSY models with quark-squark alignment, in 
which case there are additional
contributions to $\Delta M_D$ from box diagrams with gluinos and squarks
\cite{NS93}, models with a fourth quark generation in which $\Delta M_D$
gets new contributions from the $W$ and $b^\prime$ intermediate
states \cite{Babu88}, models with an $SU(2)$-singlet left-handed up-type
quark $u_L^\prime$, inducing tree-level FCNC couplings, for example in the 
form of a $Zu\bar{c}$ coupling \cite{Branco94} with implications for
the unitarity of the quark mixing matrix,
 multiscalar models with natural
flavour conservation \cite{Abbott80}, in which $\Delta M_D$ gets new
contributions from box diagrams with intermediate charged Higgs $H^\pm$ and 
quarks.
Finally, leptoquark models with light scalar leptoquarks \cite{Davidson94}
may also lead to a large value for $\Delta M_D$;
with leptoquark couplings $F_{\ell c}F_{\ell u} \geq 10^{-3}$ and leptoquark
masses $M_{LQ} \leq 2$ TeV, new contributions could be of the order of
the present experimental bounds \cite{Blaylock}.

 The present limit on $D^0$ - $\overline{D^0}$ mixing can be  
expressed in terms of the quantity $r_D \equiv (\Delta M_D/\Gamma_D)^2/
[2 +(\Delta M_D/\Gamma_D)^2]$. The decay modes of interest here are 
$D^0 \to K^+ \pi^-$ and $D^0 \to K^+ \pi^- \pi^+ \pi^-$, which can be
reached both via a doubly suppressed Cabibbo (DSC) decay and $D^0$ - 
$\overline{D^0}$ mixing. Decay time information is therefore required to
distinguish the two mechanisms.
The present experimental limit is somewhat porous, namely at $90\%$ C.L.
one has $r_D \leq 5.0 \times 10^{-3}$ in each of the two decay modes from
the E691 experiment \cite{PDG96}, assuming no interference between the
DSC- and mixing-amplitudes. As pointed out in 
\cite{Blaylock}, taking into account this interference the upper
limit is degraded, and one gets instead $r_D < 0.019$ from 
the $K^+ \pi^-$ mode and $r_D \leq 0.007$ from the $K^+ 
\pi^- \pi^+ \pi^-$ mode \cite{PDG96}. A monte carlo study to estimate
the mixing reach at HERA$(ep)$ has been done by Tsipolitis 
\cite{Tsipolitis96}, with the conclusions that a factor of 5 
improvement in $r_D$ is conceivable, given the assumed luminosity and vertex
resolution.      

 In the workshop, also experimental developments were discussed.
Jeff Appel (FNAL) gave an overview of the physics program at Fermilab 
in the next years with respect to heavy quark physics. Manfred Paulini (LBL) 
reported on the achievements of CDF and Kerstin H\"opfner (CERN) presented
a novel, radiation-hard vertex detector using scintillation fibres.

%

%%%%%%%%%%%%%%%%%%%%% REFERENCES %%%%%%%%%%%%%%%%%%%%%%%%%%%%%%%%


\vspace*{2mm}
\begin{thebibliography}{99}
%
\bibitem{CPV64}
J.H. Christenson, J.W. Cronin, V.L. Fitch, and R. Turlay,
Phys. Rev. Lett. {\bf 13} (1964) 138.

\bibitem{PDG96}
R.M. Barnett et al. (Particle Data Group), Phys. Rev. {\bf D54} (1996) 1.

\bibitem{CKM} N. Cabibbo, Phys.\ Rev.\ Lett.\ {\bf 10} (1963) 531; 
M. Kobayashi and K. Maskawa, Prog.\ Theor.\ Phys.\ {\bf 49} (1973) 652.

\bibitem{ALI96} A. Ali, preprint DESY 96-106 [hep-ph/9606324]; to appear in
the Proceedings of the XX International Nathiagali Conference on 
Physics and Contemporary Needs, Bhurban, Pakistan, June 24-July 13, 
1995 (Nova Science Publishers, New York, 1996); and in these 
proceedings.

\bibitem{BBL95}
G. Buchalla, A.J. Buras, and M.E. Lautenbacher, Preprint MPI-Ph/95-104 //
[hep-ph/9512380]; A.J. Buras, preprint TUM-HEP-255/96 [hep-ph/9609324].

\bibitem{Gibbons96} L. Gibbons (CLEO Collaboration), Invited talk at the
International Conference on High Energy Physics, Warsaw, ICHEP96 (1996).
6505.

\bibitem{EFHERA96}
R. Eichler and S. Frixione, companion summary report, these proceedings.

\bibitem{GIM}
S. Glashow, J. Iliopoulos and L. Maiani, Phys. Rev. {\bf D2} (1970) 1285.

\bibitem{Grab96}
C. Grab, these proceedings.
\bibitem{Tsipolitis96}
G. Tsipolitis, these proceedings.

\bibitem{AL96} A. Ali and D. London, preprint DESY 96-140,
UdeM-GPP-TH-96-45, [hep-ph/9607392], to appear in the {\it Proc.\ of QCD
Euroconference 96}, Montpellier, July 4-12, 1996; and in these proceedings.

\bibitem{Wolfenstein} L. Wolfenstein, Phys.\ Rev.\ Lett.\ {\bf 51} (1983)
1945.

\bibitem{superweak}
L. Wolfenstein, Phys. Rev. Lett. {\bf 13} (1964) 562.

\bibitem{misuk1} R.Misuk (HERA-B Collaboration), these proceedings.

\bibitem{claire} C.Shepherd-Themistocleous (HERA-B Collaboration),
 these proceedings.

\bibitem{misuk2} R.Misuk (HERA-B Collaboration), these proceedings.

\bibitem{Burasetal}
A.J. Buras, M. Jamin and P.H. Weisz, Nucl. Phys. {\bf B347} (1990) 491.

\bibitem{ALEPHxs} ``Combined limit on the $B_s^0$ oscillation frequency",
contributed paper by the ALEPH collaboration to the International  
Conference on High Energy Physics, Warsaw, ICHEP96 PA08-020 (1996).

\bibitem{JThom96}
J. Thom (HERA-B Collaboration), to be published.

\bibitem{CLEOrare2}
M.S. Alam et al. (CLEO Collaboration), Phys. Rev. Lett. {\bf 74} (1995) 2885.

\bibitem{CLEOrare1}
 R. Ammar et al. (CLEO Collaboration), Phys. Rev. Lett. {\bf 71} (1993) 674.


\bibitem{CLEOwarsaw}
 R. Ammar et al. (CLEO Collaboration), contributed paper to the International
Conference on High Energy Physics, Warsaw, 25 - 31 July 1996, CLEO CONF
96-05.


\bibitem{Saadi96}
 F. Saadi-L\"udemann (HERA-B Collaboration), these proceedings.


 \bibitem{InamiLim}
        T. Inami and C.S. Lim,
        Prog. Theor. Phys. {\bf 65} (1981) 297.

\bibitem{Richman96} J. Richman, Invited talk at the
International Conference on High Energy Physics, Warsaw, ICHEP96 (1996).


\bibitem{ag2}  A. Ali and C. Greub ,
               Phys. Lett. {\bf B287} (1992) 191.
\bibitem{aag96}
A. Ali, H.M. Asatrian, and C. Greub, to be published.

\bibitem{singer96}
P. Singer and D.-X. Zhang, Phys. Lett. {\bf B383} (1996) 351; H.-Y. Cheng 
and B. Tseng, Phys. Rev. {\bf D53} (1996) 1457.
%
\bibitem{abs93}
A. Ali, V.M. Braun and H. Simma, Z. Phys. {\bf C63} (1994) 437.

 
\bibitem{wyler95}
A. Khodzhamirian, G. Stoll, and D. Wyler, Phys. Lett. {\bf B358} (1995) 129.

\bibitem{ab95}
A. Ali and V.M. Braun, Phys. Lett. {\bf B359} (1995) 223.
%

\bibitem{AGM94}
 A. Ali, G. F. Giudice and T. Mannel, Z. Phys. {\bf C67} (1995) 417.

\bibitem{amm91}
        A. Ali, T. Mannel and T. Morozumi, Phys. Lett. {\bf B273} (1991) 505.

 \bibitem{AHHM96}
A.Ali, G. Hiller, L.T. Handoko, and T. Morozumi, preprint DESY 96-206,
Hiroshima report HUPD-9615 (1996) [hep-ph/9609449].

\bibitem{D0warsaw}
S. Abachi et al. (D0 Collaboration),
contributed paper  to the International
Conference on High Energy Physics, Warsaw, ICHEP96 (1996).

\bibitem{AlipietAlt}
     A. Ali and E. Pietarinen,
      Nucl. Phys. {\bf B154} (1979) 519;
     G. Altarelli et al.,
     Nucl. Phys. {\bf B208} (1982) 365.
        
\bibitem{CMW96}
P. Cho, M. Misiak, and D. Wyler, Phys. Rev. {\bf D54} (1996) 3329.
 
\bibitem{kreuzer} P.Kreuzer (HERA-B Collaboration), these proceedings

\bibitem{Beneke96}
 M. Beneke, G. Buchalla, I. Dunietz, preprint SLAC-PUB-7165 (1996)  
[hep-ph/9605259].

\bibitem{Appel94}
J.A. Appel, in \underline{Proc. of XXVII Int. Conf. on High Energy Physics},
Glasgow, 20 - 27 July 1994 (IOP Publishing, 
1995); eds.: P.G. Bussey and I.P. Knowles.

\bibitem{Egli91}
S. Egli et al., in \underline{Physics at HERA}, Vol. 2 (1992) 770,
 eds.: W. Buchm\"uller and G. Ingelman.

\bibitem{Georgi}
J.F. Donoghue et al., Phys. Rev. {\bf D33} (1986) 179; H. Georgi, Phys. Lett.
{\bf B297} (1992) 353; T. Ohl, G. Ricciardi and E.H. Simmons,
Nucl. Phys. {\bf 403} (1993) 605. See also L. Wolfenstein, Phys. Lett. 
{\bf B164} (1985) 170.

\bibitem{Hewett} 
For a recent review, see J.L. Hewett, T. Takeuchi and S. Thomas, preprint
 SLAC-PUB-7088 (1996), [hep-ph/9603391].

\bibitem{NS93}
Y. Nir and N. Seiberg, Phys. Lett. {\bf B309} (1993) 337;
M. Leurer, Y. Nir and N. Seiberg, Nucl. Phys. {\bf B420} (1994) 468.

\bibitem{Babu88}
K.S. Babu, X.-G. He, X.-Q. Li and S. Pakvasa, Phys. Lett. {\bf B205}
(1988) 540.

\bibitem{Branco94}
G.C. Branco, P.A. Parada and M.N. Rebelo, Phys. Rev. {D52} (1995) 4217.

\bibitem{Abbott80}
L.F. Abbott, P. Sikivie and M.B. Wise, Phys. Rev. {\bf D21} (1980) 1393;
V. Barger, J.L. Hewett and R.J.N. Phillips, Phys. Rev. {\bf D41} (1990) 3421;
Y. Grossman, Nucl. Phys. {\bf B246} (1994) 355.

\bibitem{Davidson94}
W. Buchm\"uller and D. Wyler, Phys. Lett. {\bf B177} (1986) 377;
S. Davidson, D. Bailey and B.A. Campbell, Z. Phys. {\bf C61} (1994) 613;
M. Leurer, Phys. Rev. Lett. {\bf 71} (1993) 1324; Phys. Rev. {\bf D48}
(1994) 333.

 \bibitem{Blaylock}
G. Blaylock, A. Seiden and Y. Nir, Phys. Lett. {\bf B355} (1995) 555.

\end{thebibliography}
\end{document}